\documentclass[runningheads]{llncs}
\usepackage{url}
\usepackage{cite}
\usepackage{algorithmic}
\usepackage{algorithm}
\usepackage{amsmath}
\usepackage{array}
\usepackage{graphicx}
\usepackage{multicol}
\usepackage{multirow}
\usepackage{subfig}

\usepackage[shadow,backgroundcolor=orange!40]{todonotes}

\newcommand{\ie}{i.\,e.}
\newcommand{\eg}{e.\,g.}

\newcolumntype{L}[1]{>{\raggedright\arraybackslash}p{#1}} 
\newcolumntype{C}[1]{>{\centering\arraybackslash}p{#1}}   
\newcolumntype{R}[1]{>{\raggedleft\arraybackslash}p{#1}}  

\newcommand{\LINEIF}[2]{\STATE\algorithmicif\ {#1}\ \algorithmicthen\ {#2} \algorithmicend\ \algorithmicif}
\newcommand{\LINEFORALL}[2]{\STATE\algorithmicforall\ {#1}\ \algorithmicdo\ {#2} \algorithmicend\ \algorithmicfor}

\begin{document}
\title{Little Boxes: A Dynamic Optimization Approach for Enhanced Cloud Infrastructures}
\titlerunning{Little Boxes: Dynamic Optimization for Enhanced Cloud Infrastructures}
\author{Ronny Hans\inst{1} \and
Bj\"orn Richerzhagen\inst{1} \and
Amr Rizk\inst{1} \and
Ulrich Lampe\inst{1} \and
Ralf Steinmetz\inst{1} \and
Sabrina Klos (n\'{e}e M\"{u}ller)\inst{2} \and
Anja Klein\inst{2}}
\authorrunning{Ronny Hans et al.}
%
%
\institute{Technische Universit\"at Darmstadt, Multimedia Communications Lab (KOM), \\ Rundeturmstr. 10,
64283 Darmstadt, Germany \\ \email{Ronny.Hans@KOM.tu-darmstadt.de}\\
\and
Technische Universit\"at Darmstadt, Communications Engineering Lab, \\ Merckstra{\ss}e 25, 64283 Darmstadt, Germany }
\maketitle              
\begin{abstract}
The increasing demand for diverse, mobile applications with various degrees of Quality of Service requirements meets the increasing elasticity of on-demand resource provisioning in virtualized cloud computing infrastructures.
This paper provides a dynamic optimization approach for enhanced cloud infrastructures, based on the concept of cloudlets, which are located at hotspot areas throughout a metropolitan area.
In conjunction, we consider classical remote data centers that are rigid with respect to QoS but provide nearly abundant computation resources.
Given fluctuating user demands, we optimize the cloudlet placement over a finite time horizon from a cloud infrastructure provider's perspective.
By the means of a custom tailed heuristic approach, we are able to reduce the computational effort compared to the exact approach by at least three orders of magnitude, while maintaining a high solution quality with a moderate cost increase of 5.8\% or less.

\end{abstract}
\section{Introduction} \label{sec:introduction}

\noindent Over the last decade, the development of Information Technology (IT) has been shaped by different trends.
One of these trends is cloud computing, which started as a paradigm for monetizing surplus IT resources to become a cornerstone paradigm in resource provisioning for business as well as private customers.
In addition to these trend, we observed another major trend of increasing dissemination of mobile devices over the past few years.
Omnipresent smartphones are heavily used today to consume multimedia services, communicate, and play massive real-time online games.

Combining these two trends together, i.e., \emph{(i)} the demand for more diverse services -- especially given device mobility -- together with \emph{(ii)} the elastic on-demand service (resource) provisioning of the cloud computing paradigm, we arrive at the mobile cloud computing paradigm.
This paradigm imposes many new challenges, specifically regarding the Quality of Service (QoS) requirements of mobile services.
Strict QoS requirements while providing multimedia services stand in contrast to the usual concentration of computational resources in a small number of large, centralized cloud data centers.
To reduce the latency between data centers and users, research showed that a higher service quality can be achieved with an increased number of data centers.
This obviously causes immense additional costs and oppose the \emph{economies of scale} advantage of cloud computing \cite{Choy2012Brewing,Goiri2011Intelligent}.

%
Mobile devices using LTE networks suffer from higher latency \cite{Sommers2012Cell} and high energy consumption \cite{huang2012close}.
Such problems can be addressed by utilizing (miniature) data centers or computation resources in proximity to the user.
In the best case, such resources are accessible via Wi-Fi and offer interfaces to offload the computation of intensive tasks.
These resources at the edge of the network are referred to as \emph{cloudlets} \cite{Satyanarayanan2009Case}.
In the work at hand, we investigate a \emph{cost-efficient} and \emph{QoS-aware} placement of cloudlet resources using a time dynamic, multi-period optimization model.
The remainder of the paper is structured as follows: In Section~\ref{sec:problem_statement}, we provide a problem description from a provider's perspective.
Section~\ref{sec:optimization} contains our model and the solution approach.
Section~\ref{sec:Evaluation} provides a quantitative evaluation of the results.
Subsequently, in Section~\ref{sec:relatedwork}, we provide an overview of relevant related work on cloudlets and edge resource placements, before concluding the paper in Section~\ref{sec:conclusion}.
%
%
\section{Problem Statement}\label{sec:problem_statement}
In this work, we assume the role of a cloud infrastructure provider that aims to provide resources for higher layer application service providers.
We assume that the provider owns the cloud infrastructure at hand and, thus, has free disposure over all of its resources.
For premium services with rigid QoS constraints, the provider aims to augment his infrastructure using cloudlets within a metropolitan area.
Therefore, we consider stationary cloudlets with permanently installed hardware, which are connected to the same Local Area Network (LAN), \ie, Wi-Fi, as the users \cite{Satyanarayanan2009Case, Verbelen2012Cloudlets}. Hence, the users benefit from a low propagation delay and a high bandwidth. As deployment method, we assume a top-down approach, where the provider owns and offers cloudlets and, hence, bears the entrepreneurial risk \cite{Satyanarayanan2009Case}. We consider cloudlet locations at existing restaurants or cafes (\eg, Starbucks stores) in Manhattan. Obviously, such deployments require contractual agreements. Since we are focusing on the optimization approaches, the underlying business models are out of scope for this paper.

\begin{figure}[b]
\centering
\includegraphics[width=.5\textwidth]{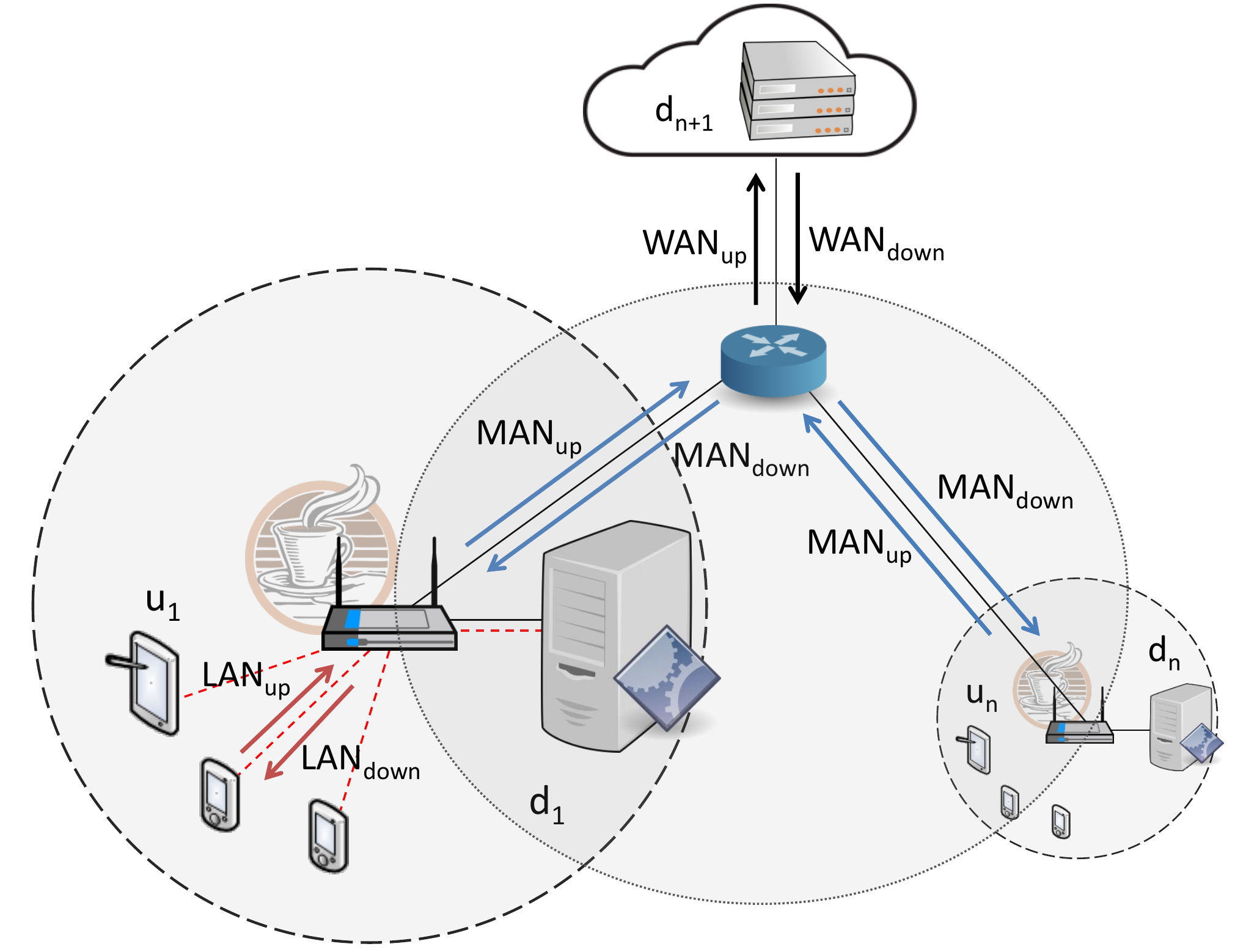}
\caption{Integration of cloudlets within a network topology}
\label{fig:dyn_cloudlet_man}
\vspace{-10pt}
\end{figure}

In the following, we aggregate all users covered by a local Wi-Fi into a user cluster with a defined demand for services.
Naturally, this user demand is fluctuating over time.
As depicted in Figure \ref{fig:dyn_cloudlet_man}, a user cluster comprises different types of network connections.

First, a hard-wired LAN connects the Wi-Fi hotspot, a possibly installed cloudlet, and the router to communicate to external remote resources.
Second, Wi-Fi connections that connect the mobile devices to the Wi-Fi hotspot.
Since we are assuming a higher bandwidth on the wired LAN compared to the wireless Wi-Fi hotspot, we do not consider the LAN as a limiting factor.

The third network component connects a user cluster to a central router within the Metropolitan Area Network (MAN) and hence, to other user clusters, cloudlets, and remote data centers.
Figure~\ref{fig:dyn_cloudlet_man} shows the basic structure of a cloudlet, the networks, and the connection to a remote cloud data center.
The provider may place cloudlets and the corresponding resources at different locations.
When putting a new location into service, fixed infrastructure cost will arise. 
Each cloudlet can be equipped with a number of servers up to an upper capacity bound.
The capacity is restricted by limited physical space, limited feasibility for cooling, or restrictions regarding the overall energy consumption.
For each deployed server, fixed hardware costs occur.
Furthermore, for each resource unit variable costs arise, \eg, for electricity and cooling.
Since such costs may fluctuate over time, \eg, due to varying energy prices, a provider needs to consider a planning time horizon that is captured here through multiple time periods.
If a resource migration, \eg, in form of VM migration, is required, migration costs arise.
We assume that these costs are independent of the type of cloudlet or the distance between the cloudlets.
In real world scenarios, service migrations can be time aligned with data transfer.
Therefore, we consider different migration costs depending on the service class (cf. Table~\ref{tab:service_req}).
In our model, penalty costs arise if a specific user demand cannot be fulfilled.

Data centers provide different QoS guarantees with respect to each user cluster, \ie, with respect to the end-to-end latency that depends on the distance between the data center and the user cluster.
Therefore, a provider needs to differentiate between the different types of data centers for service placement, \ie, local cloudlets and remote data centers.
The latter one generally possesses a higher latency.

By the means of the provided infrastructure, users access various services.
We distinguish between three different service classes, whereby each class possesses specific QoS requirements:
\emph{(i)} Cloud services that can be easily used via a cellular network, \ie, services with low QoS requirements regarding latency and bandwidth, for example messaging tools.
\emph{(ii)} Cloud services that can be easily used via broadband internet, \ie, services with high bandwidth requirement, but not necessarily realtime constraints, such as on-demand video streaming.
\emph{(iii)} Cloud services with high computational effort, realtime constraints, and high bandwidth requirements, \eg, cloud gaming.

The first class of services plays a minor role in our scenario, since cloudlets only offer marginal additional benefits to such services.
Nevertheless, these services can by provided by cloudlets if free capacities are available.
For the second class of services, cloudlets increase the users' quality of experience through a high bandwidth to demanded content.
For the third class of services, we note that cloudlets are required to ensure appropriate quality of service guarantees.

The purpose of this optimization, which is based on a provider's perspective, is to place resources in data centers and take decisions regarding the required capacity while providing QoS guarantees.
Thereby, the goal is a minimization of the overall provisioning costs.
In the following, we refer to this problem as \emph{Dynamic Cloudlet Placement and Selection Problem (DCPSP)}.
%
%
\section{Optimization Approach}\label{sec:optimization}
Next, we present a Mixed Integer Linear Program (MILP) formulation for the dynamic cloudlet placement and selection problem.
Subsequently, in order to solve the problem in an efficient manner, we provide a heuristic solution approach, which is described in detail in Section~\ref{subsec:heuristic}.
%
%
\subsection{Exact Approach}\label{subsec:optimization_exa}
To provide a mathematical model for the \emph{DCPSP}, we introduce the formal notation in Table~\ref{table:notations}.
The objective is the minimization of the total monetary cost associated with the cloudlet placement and selection. 
%
%
\subsubsection{Optimization Goal}
The objective function aiming to minimize the total costs is given in Eq.~\ref{eq:objective}.
These costs are split into fixed infrastructure cost, variable operating cost, variable reservation cost, penalty cost, migration cost, and fixed hardware cost.
{\small
\begin{align}
\label{eq:objective}
& {\min~ }C = \nonumber \\
& \sum_{\substack{{\lambda=1..\Lambda}}} x_{d_\lambda} \times C^{fix}_{d_\lambda} + \sum_{o=1..O}(\sum_{\substack{{\lambda=1..\Lambda} \\  {\mu=1..M} \\ {\nu=1..N}}} y_{d_\lambda,u_\mu,s_\nu,t_o} \times C^{op}_{d_\lambda,t_o} + \sum_{\substack{{\mu=1..M} \\ \nu=1..N}} y^{pen}_{u_\mu,s_\nu,t_o} \times C^{pen}_{u_\mu,s_\nu}) \nonumber \\
& + \sum_{o=2}^{O}\sum_{\substack{{\lambda=1..\Lambda} \\ {\mu=1..M} \\ \nu=1..N}}y^{mig}_{d_\lambda,u_\mu,s_\nu,t_o} \times C^{mig}_{s_\nu} + \sum_{{\lambda=1..\Lambda}} z_{d_\lambda} \times C^{hw}_{d_\lambda}
\end{align}
}
The first summand represents the fixed infrastructure cost that depends on the selected data centers represented by the decision variable $x_{d_\lambda}$ and the corresponding value for the individual fixed cost $C^{fix}_{d_\lambda}$.
Such resource-agnostic cost occurs once for each planning period when a data center is placed.
The second part of the term summarizes to the variable operational costs $C^{op}_{d_\lambda,t_o}$ that are caused by the provided resource units $y_{d_\lambda,u_\mu,s_\nu,t_o}$.
The operational costs depend on the selected data center and may well vary over time.
The third summand refers to capacities requested by a user cluster $u_\mu$ that are unfulfilled by the selected data centers.
These capacities, $y^{pen}_{u_\mu,s_\nu,t_o}$, cause penalty cost $C^{pen}_{u_\mu,s_\nu}$.
The migration cost is expressed in the fourth summand.
Such migration cost $C^{mig}_{u_\mu,s_\nu}$ includes the data transfer cost from one data center to another.
Since we assume that launching a new service does not cause migration cost, such cost only occurs from the second time period on.

Eq.~\ref{eq:migrationscost} expresses the number of resource units to be migrated.
To calculate the total amount, we distinguish two different cases:
\emph{(i)} The amount of resources that is provided to a specific user cluster $u_\mu$ w.r.t. a specific service is either constant or increases between two subsequent time periods, while the resource share provided by specific data center decreases.
\emph{(ii)} the aggregated amount of resources provided to a specific user cluster $u_\mu$ w.r.t. a specific service decreases between to time slots, while the resource share provided by a specific data center increases.
To model and implement the optimization problem, this case differentiation requires a standard transformation into a linear equation system.
However, due to space restrictions, this transformation is not part of the work at hand.
{\small
\begin{align}
\label{eq:migrationscost}
& y^{mig}_{d_\lambda,u_\mu,s_\nu,t_o} =
	\begin{cases}
		& y_{d_\lambda,u_\mu,s_\nu,t_{o-1}} - y_{d_\lambda,u_\mu,s_\nu,t_o} \quad {\text{if} } \nonumber \\
		&	\quad \sum_{{\alpha=1..\Lambda}} y_{d_\alpha,u_\mu,s_\nu,t_o} \geq \sum_{{\alpha=1..\Lambda}} y_{d_\alpha,u_\mu,s_\nu,t_{o-1}} \nonumber \\
		& \quad \wedge ~y_{d_\lambda,u_\mu,s_\nu,t_o} \leq y_{d_\lambda,u_\mu,s_\nu,t_{o-1}}\\
		& y_{d_\lambda,u_\mu,s_\nu,t_o} - y_{d_\lambda,u_\mu,s_\nu,t_{o-1}} \quad {\text{if} } \nonumber \\
		&	\quad \sum_{{\alpha=1..\Lambda}} y_{d_\alpha,u_\mu,s_\nu,t_o} < \sum_{{\alpha=1..\Lambda}} y_{d_a,u_\mu,s_\nu,t_{o-1}} \nonumber \\
		& \quad \wedge ~y_{d_\lambda,u_\mu,s_\nu,t_o} > y_{d_\lambda,u_\mu,s_\nu,t_{o-1}}\\
		& 0 \quad \text{else}
	\end{cases} \nonumber\\
&	\forall d_\lambda \in D, \forall u_\mu \in U, \forall s_\nu \in S, \forall t_o \in T
\end{align}
}
Note that the last summand in Eq.~\ref{eq:objective} refers to the provided hardware units $z_{d_\lambda}$ in each data center.
Providing servers leads to hardware cost $C^{hw}_{d_\lambda}$.
\begin{table}[!h]
\caption{Formal notations \label{table:notations}}
\footnotesize
\renewcommand{\arraystretch}{1.35}
\begin{center}
\begin{tabular} {C{2cm}|L{10cm}}
\hline\hline \\[-1em]
\textbf{Symbol} & \textbf{Description} \\[0.25em] \hline
$d_\lambda$ & represents a specific data center and encompasses cloud data centers and cloudlets \\
$u_\mu $ & represents a specific user cluser \\
$s_\nu$ & represents a specific service \\
$q_\xi$ & represents a specific QoS attribute \\
$t_o $ & represents a specific time slot within the planning period \\
$V_{u_\mu,s_\nu,t_o}$ & service demand of user $u_\mu$ for service $s_\nu$ at time $t_o$ \\
$K^{min}_{d_\lambda}$ & minimal capacity of data center $d_\lambda$ \\
$K^{max}_{d_\lambda}$ & maximal capacity of data center $d_\lambda$ \\
$K^{LAN_{down}}_{u_\mu}$ & LAN downlink capacity of user cluster $u_\mu$ \\
$K^{LAN_{up}}_{u_\mu}$ & LAN uplink capacity of user cluster $u_\mu$ \\
$K^{MAN_{down}}_{u_\mu}$ & WAN downlink capacity of user cluster $u_\mu$ \\
$K^{MAN_{up}}_{u_\mu}$ & WAN uplink capacity of user cluster $u_\mu$ \\
$C^{fix}_{d_\lambda}$ & fixed cost of selecting data center $d_\lambda$ \\
$C^{hw}_{d_\lambda}$ & fixed costs for buying or leasing hardware for data center $d_\lambda$ \\
$C^{op}_{d_\lambda,t_o}$ & variable cost for operating one resource unit for one time unit in data center $d_\lambda$ at time $t_o$ \\
$C^{mig}_{s_\nu}$ & migration cost for moving service $s_\nu$ from one data center to another between two subsequent time periods $t$ and $t+1$ \\
$C^{pen}_{u_\mu,s_\nu}$ & penalty cost per service unit not provided to user $u_\mu$ w.r.t. service $s_\nu$ \\
$Q^{gua}_{d_\lambda,u_\mu,q_\xi}$ & QoS guarantee of data center $d_i$ w.r.t. user $u_j$ for QoS attribute $q_\xi$ \\
$Q^{req}_{u_\mu,s_\nu,q_\xi}$ & QoS requirement of user $u_i$ w.r.t. service $s_\nu$ for QoS attribute $q_\xi$ \\
$L^{down}_{s_\nu}$ & required downstream capacity for service $s_\nu$ \\
$L^{up}_{s_\nu}$ & required upstream capacity for service $s_\nu$ \\
$x_{d_\lambda}$ & variable $\in \{0,1\}$ indicates whether a data center $d_\lambda$ will be used or not\\
$y_{d_\lambda,u_\mu,s_\nu,t_o}$ & number of resources a data center $d_\lambda$ provides to a user cluster $u_\mu$ regarding a service $s_\nu$ in time period $t_o$ \\
$y^{mig}_{d_\lambda,u_\mu,s_\nu,t_o}$ & number of resources that are migrated from one to another data center in between the time periods $t_{o-1}$ and $t_o$ \\
$y^{pen}_{u_\mu,s_\nu,t_o}$ & demand that is not satisfied by the provider and that will cause penalty costs \\
$z_{d_\lambda}$ & number of hardware resource units provided within a data center $d_\lambda$ \\
\hline\hline
\end{tabular}
\end{center}
\vspace{-10pt}
\end{table}
%
%
%
\subsubsection{Constraints}
In the following, we present the required constraints to ensure a valid solution of this optimization problem.
The first constraint in Eq.~\ref{eq:ensure_demand_description_stat} concerns the user cluster demand $V_{u_\mu,s_\nu,t_o}$. 
Since a provider has the choice either to fulfill the demand or cause a penalty, the summation of provided and unfulfilled capacities must be equal or greater to the resource demand of all user clusters for all services at each point in time.
%
{\small
\begin{align}
\label{eq:ensure_demand_description_stat}
&	y^{pen}_{u_\mu,s_\nu,t_o} + \sum_{\lambda=1..\Lambda} y_{d_\lambda,u_\mu,s_\nu,t_o} \geq V_{u_\mu,s_\nu,t_o} \quad \forall u_\mu \in U, \forall s_\nu \in S, \forall t_o \in T
\end{align}
}
The available data center resources are limited by a maximal capacity constraint $K^{max}_{d_\lambda}$, \eg, by the available space or cooling.
Further, we consider a lower capacity bound $K^{min}_{d_\lambda}$ reflecting the economic necessity of a cost-efficient operation of data centers.
As cloudlets can be established with few hardware resources, \eg, a single server, this bound could also be set to zero.
These conditions determine the number of hardware resources $z_{d_\lambda}$ that can be installed within a data center $d_\lambda$ (cf. Eq.~\ref{eq:link_yz_description} and Eq.~\ref{eq:cap_held_description}).
{\small
\begin{align}
\label{eq:link_yz_description}
\sum_{\substack{{m=1..n} \\ {\nu=1..N}}} y_{d_\lambda,u_\mu,s_\nu,t_o} \leq z_{d_\lambda} \quad \forall d_\lambda \in D, \forall t_o \in T
\end{align}
\begin{align}
\label{eq:cap_held_description}
z_{d_\lambda} \leq x_{d_\lambda} \times K^{max}_{d_\lambda} \quad \forall d_\lambda \in D,
z_{d_\lambda} \geq x_{d_\lambda} \times K^{min}_{d_\lambda} \quad \forall d_\lambda \in D
\end{align}
}
The adherence to QoS requirements is expressed by the binary variable $p_{d_\lambda,u_\mu,s_\nu}$.
If all QoS guarantees $Q^{gua}_{d_\lambda,u_\mu,q_\xi}$ are fulfilled, the variable is set to \emph{one} (cf. Eq.~\ref{eq:qos_satisfied_case_description}).
Otherwise, a data center cannot provide any resources (cf. Eq.~\ref{eq:qos_satisfied_description}).
%
{\small
\begin{align}
\label{eq:qos_satisfied_case_description}
p_{d_\lambda,u_\mu,s_\nu} =
	\begin{cases}
		& 1 \quad \text{if } Q^{gua}_{d_\lambda,u_\mu,q_\xi} \geq Q^{req}_{u_\mu,s_\nu,q_\xi} \forall q_\xi \in Q \\
		& 0 \quad \text{else}
	\end{cases}
\end{align}
\begin{align}
\label{eq:qos_satisfied_description}
& y_{d_\lambda,u_\mu,s_\nu,t_o} \leq p_{d_\lambda,u_\mu,s_\nu} \times K^{max}_{d_\lambda} \quad \forall d_\lambda \in D, \forall u_\mu \in U, \forall s_\nu \in S, \forall t_o \in T
\end{align}
}
As described earlier, each user cluster is connected to two types of networks, a LAN, \ie, Wi-Fi, and a MAN that connects the different user clusters with each other and to remote cloud data centers.
All services that are consumed require a specific average amount of bandwidth.
Note that the required bandwidth most be lower or equal than the available bandwidth.
Since services may have different requirements regarding download and upload capacities, we differentiate between these two (cf. Eq.~\ref{eq:ConstraintLinkLanDown_description} and~\ref{eq:ConstraintLinkLanUp_description}).
{\small
\begin{align}
\label{eq:ConstraintLinkLanDown_description}
& \sum_{\substack{{\lambda=1..\Lambda}}} \sum_{\nu=1..N} y_{d_\lambda,u_\mu,s_\nu,t_o} \times L^{down}_{s_\nu} \leq K^{LAN_{down}}_{u_\mu} \nonumber \\
&\forall u_\mu \in U, \forall s_\nu \in S, \forall t_o \in T
\end{align}
\begin{align}
\label{eq:ConstraintLinkLanUp_description}
& \sum_{\substack{{\lambda=1..\Lambda}}} \sum_{\nu=1..N} y_{d_\lambda,u_\mu,s_\nu,t_o} \times L^{up}_{s_\nu} \leq K^{LAN_{up}}_{u_\mu} \nonumber \\
& \forall u_\mu \in U, \forall s_\nu \in S, \forall t_o \in T
\end{align}
}
The MAN connection is required to provide services from remote resources to a local user cluster, and may be necessary to provide services from a local cloudlet to remote users.
For services that are provided by the \emph{local} cloudlet and consumed by the \emph{local} users, no MAN capacities are required at all.
Eq.~\ref{eq:ConstraintLinkMANDown_description} and Eq.~\ref{eq:ConstraintLinkMANUp_description} represent the corresponding constraints.
Further, we differentiate between download and upload capacities to take specific service requirements and network characteristics into account.
{\small
\begin{align}
\label{eq:ConstraintLinkMANDown_description}
& \sum_{\substack{{\lambda=1..\Lambda} \\ \lambda \ne \alpha}} \sum_{\nu=1..N} y_{d_\lambda,u_\alpha,s_\nu,t_o} \times L^{down}_{s_\nu} + \sum_{\substack{{\mu=1..M} \\ \mu \ne \alpha}} \sum_{\nu=1..N} y_{d_\alpha,u_\mu,s_\nu,t_o} \times L^{up}_{s_\nu}
\quad \leq K^{MAN_{down}}_{u_\alpha} \nonumber \\
& \forall d_\alpha \in D, \forall u_\alpha \in U, \forall s_\nu \in S, \forall t_o \in T
\end{align}
\begin{align}
\label{eq:ConstraintLinkMANUp_description}
& \sum_{\substack{{\lambda=1..\Lambda} \\ \lambda \ne \alpha}} \sum_{\nu=1..N} y_{d_\lambda,u_\alpha,s_\nu,t_o} \times L^{up}_{s_\nu} + \sum_{\substack{{\mu=1..M} \\ \mu \ne \alpha}} \sum_{\nu=1..N} y_{d_a,u_\mu,s_\nu,t_o} \times L^{down}_{s_\nu}
\quad \leq K^{MAN_{up}}_{u_\alpha} \nonumber \\
& \forall d_\alpha \in D, \forall u_\alpha \in U, \forall s_\nu \in S, \forall t_o \in T
\end{align}
}
The presented optimization problem constitutes a Mixed Integer Program (MIP) and is NP-hard.
Hence, no known algorithms exists that is capable to solve any corresponding problem instances in polynomial time.
In the following, we describe a heuristic solution approach to obtain solutions to this problem with reasonable effort.
%
%
\subsection{Heuristic Approach}\label{subsec:heuristic}
To find solutions to the DCPSP in reasonable time and in an appropriate solution quality we investigate in different heuristic solution strategies.
Therefore, we adapted our static approach from \cite{Hans2015Setting} to this dynamic optimization problem.
We improve this single period approach by enhancing the algorithm to optimize across multiple periods and by including network capacities and migration cost.
We also evaluate our approach using multiple services classes with different non-functional requirements.

The pseudo code of our approach is illustrated in Algorithm~\ref{alg:heu_dcpsp}.
The additional required variables are explained in Table~\ref{table:notations_heu}.
With the goal in mind to minimize the total cost, we implement and evaluate two different strategies:
\emph{(i)} Cover as much demand as possible and \emph{(ii)} limit the provided resources to avoid an over provisioning.
The first strategy, DCPSP-HEU1.KOM, can freely dispose over all available resources within the boundaries given by the earlier described constraints.
We assume this strategy causes high fixed cost but minimizes the penalty cost.
Since we aim to satisfy as much demand as possible, this strategy also implies a high user satisfaction.
With the second strategy, DCPSP-HEU2.KOM, we aim to prevent a resources provisioning for single peak load.
Providing resources for such scarce events might increase the overall fixed cost.
Thus, this strategy avoids over-provisioning and, therefore, trades penalty costs for fixed costs.
\begin{table}[!h]
\caption{Additional formal notations for the heuristic approach \label{table:notations_heu}}
\footnotesize
\renewcommand{\arraystretch}{1.35}
\begin{center}
\begin{tabular} {C{2cm}|L{10cm}}
\hline\hline \\[-1em]
\textbf{Symbol} & \textbf{Description} \\[0.25em] \hline
$D^{\text{per}}_{u_\mu,s_\nu,t_o}$ & List of permitted data centers depend on QoS requirements and guarantees \\
$U^{\text{res}}_s$ & List of residual user clusters and their requested services \\
$V^{\text{res}}_{u_\mu,s_\nu}$ & Residual demand w.r.t. user cluster and data center \\
$V^{\text{ass}}_{d_\lambda,u_\mu,s_\nu,t_o}$ & Already assigned demand w.r.t. user cluster, data center, and time period \\
$K^{\text{res}}_{d_\lambda}$ & Residual (unused) capacities of the data centers \\
$N^{\text{res}}_{u_\mu}$ & List of residual network capacities of a user cluster (incl. LAN, MAN, down and up link) \\
$l_{LAN}$ & Lot size of a LAN for an additional quantity of services \\
$l_{MAN}$ & Lot size of a MAN for an additional quantity of services \\
$k^{max}$ & Maximal number of provided capacities by cloudlets (only required for second strategy) \\
$k^{max}_{count}$ & Counter for capacities provided by cloudlets (only required for second strategy) \\
\hline\hline
\end{tabular}
\end{center}
\end{table}
In our model, we assume that services with the highest QoS constraints cause the highest migration and penalty cost.
Hence, these services require a higher priority for assignment.
To minimize these costs, both strategies take the previously made assignment decisions into account (cf. Line~\ref{heu_dcpsp_selection:line:end_trans1} to \ref{heu_dcpsp_selection:line:end_trans2}) and retain decisions from the previous period.

For the unassigned demand, we use an iterative process, which starts at Line~\ref{heu_dcpsp_selection:line:while}.
The process is repeated until the entire service demand is assigned or no more appropriate resources are available.
Depending on the priority, we select a user cluster and a service to be assigned (cf. Line~\ref{heu_dcpsp_selection:line:select_user}).
If there are free capacities within the MAN link of a user cluster, we include all permitted data centers and select the most appropriate one (cf. Line~\ref{heu_dcpsp_selection:line:select_datacenter}).
Otherwise, we check if a local cloudlet is available for this particular user cluster and use these resources (cf. Line~\ref{heu_dcpsp_selection:line:useCloudlet}).
To decide what quantity of resources are assigned in the current iteration, we calculate the lot size depending on the residual demand, resource capacities, and network capacities (cf. Line~\ref{heu_dcpsp_selection:line:assignment}).
The assigned resources are stored in the list $V^{\text{ass}}_{d_\lambda,u_\mu,s_\nu,t_o}$ (cf. Line~\ref{heu_dcpsp_selection:line:assign}).
Finally, we adjust the list storing data center and user cluster information.
If all resources of a data center are assigned, we remove it from the list of permitted data centers (cf. Line~\ref{heu_dcpsp_selection:line:kill_datacenter}).
Further, we remove a pair user cluster and service form the corresponding list if at least one of the following conditions are true: \emph{(i)} The entire demand is covered, \emph{(ii)} no appropriate resources are provided, or \emph{(iii)} no more network capacities are available (cf. Line~\ref{heu_dcpsp_selection:line:kill_usercluster1}).
Finally, after finishing this process, we calculate the overall cost caused by the assignment, migration, and penalties.
In the subsequent section, we compare the solution obtained through this heuristic approach to the optimal one.
\begin{algorithm}
\small
	\begin{algorithmic}[1]
		\REQUIRE
		\FORALL{$t_o$}
				\STATE{// \textit{Initilize Lists}}
				\STATE $D^{\text{per}}_{u_\mu,s_\nu,t_o} \leftarrow InitDataCenters()$ \label{heu_dcpsp_selection:line:permit_dc}
				\STATE $V^{\text{res}}_{u_\mu,s_\nu} \leftarrow InitResidualDemand()$ \label{heu_dcpsp_initialize:line:res_dem}
				\STATE $K^{\text{res}}_{d_\lambda} \leftarrow InitResidualCapacity()$ \label{heu_dcpsp_selection:line:reduce_dem}
				\STATE $N^{res}_{u_\mu} \leftarrow InitResidualNetworkCapacity()$
				\STATE {// \textit{Transfer previous assignments}} \label{heu_dcpsp_selection:line:start_trans}
				\FORALL{$V^{\text{ass}}_{u_\mu,s_\nu,t_{o-1}}$} \label{heu_dcpsp_selection:line:end_trans1}
						\IF{$\nu = 1 ~or~ \nu = 2$}
							\STATE $y_{d_\lambda,u_\mu,s_\nu} \leftarrow min(V^{\text{ass}}_{d_\lambda,u_\mu,s_\nu,t_{o-1}},V^{\text{res}}_{u_\mu,s_\nu})$ \label{heu_dcpsp_selection:line:assignment1}
							\STATE $V^{\text{ass}}_{d_\lambda,u_\mu,s_\nu,t_o} \leftarrow y_{d_\lambda,u_\mu,s_\nu}$
							\STATE $V^{\text{res}}_{u_\mu,s_\nu} \leftarrow V^{\text{res}}_{u_\mu,s_\nu} - y_{d_\lambda,u_\mu,s_\nu}$
							\STATE $K^{\text{res}}_{d_\lambda} \leftarrow K^{\text{res}}_{d_\lambda} - y_{d_\lambda,u_\mu,s_\nu}$ \label{heu_dcpsp_selection:line:reduce_dem1}
							\STATE $N^{res}_{u_\mu} \leftarrow CalcResidualNetworkCap(N^{res}_{u_\mu},y_{d_\lambda,u_\mu,s_\nu})$
						\ENDIF	
				\ENDFOR \label{heu_dcpsp_selection:line:end_trans2}
					\STATE {// \textit{Assign open demand}}
					\WHILE{$|U^{\text{res}}_s| > 0 ~or~ k^{max}_{count}< k^{max}$} \label{heu_dcpsp_selection:line:while}
					\STATE {// \textit{Choose element for demand}}
					\STATE $u_\mu,s_\nu \leftarrow SelectServiceDemand(V^{\text{res}}_{u_\mu,s_\nu})$ \label{heu_dcpsp_selection:line:select_user}					
					\STATE {// \textit{Choose element for supply}}
					\STATE $l_{LAN} \leftarrow CalcLotSizeLAN(N^{res}_{u_\mu},s_\nu)$
					\STATE $l_{MAN} \leftarrow CalcLotSizeMAN(N^{res}_{u_\mu},s_\nu)$
					\IF{$l_{MAN} >= 1$} %
						\STATE $d_\lambda \leftarrow SelectDataCenter(D^{\text{per}}_{u_\mu,s_\nu})$ \label{heu_dcpsp_selection:line:select_datacenter}
					\ELSIF{$ExistLocalCloudlet$} %
						\STATE $d_\lambda \leftarrow SelectLocalCloudlet(D^{\text{per}}_{u_\mu,s_\nu})$\label{heu_dcpsp_selection:line:useCloudlet}
					\ENDIF
					\STATE {// \textit{Reduce residual quantities and save assignment}}
					\STATE $y_{d_\lambda,u_\mu,s_\nu} \leftarrow CalcLotSize(K^{\text{res}}_{d_\lambda},V^{\text{res}}_{u_\mu,s_\nu}, N^{res}_{u_\mu})$ \label{heu_dcpsp_selection:line:assignment}
					\STATE $V^{\text{ass}}_{d_\lambda,u_\mu,s_\nu,t_o} \leftarrow y_{d_\lambda,u_\mu,s_\nu}$ \label{heu_dcpsp_selection:line:assign}
					\STATE $V^{\text{res}}_{u_\mu,s_\nu} \leftarrow V^{\text{res}}_{u_\mu,s_\nu} - y_{d_\lambda,u_\mu,s_\nu}$ \label{heu_dcpsp_selection:line:reduce_cap}
					\STATE $K^{\text{res}}_{d_\lambda} \leftarrow K^{\text{res}}_{d_\lambda} - y_{d_\lambda,u_\mu,s_\nu}$ \label{heu_dcpsp_selection:line:reduce_dem2}
					\STATE $N^{res}_{u_\mu} \leftarrow CalcResidualNetworkCap(N^{res}_{u_\mu},y_{d_\lambda,u_\mu,s_\nu})$
					\LINEIF{$(d_\lambda ~isCloudlet)$}{$k^{max}_{count} += y_{d_\lambda,u_\mu,s_\nu}$}
					\STATE {// \textit{Add and remove elements from various lists}}
					\STATE $D^{\text{open}} \leftarrow D^{\text{open}} \cup \{d_\lambda\}$ \label{heu_dcpsp_selection:line:store_open_datacenters}
					\IF{$K^{\text{res}}_{d_\lambda} = 0$}
						\LINEFORALL{$\text{\{}u_\mu,s_\nu\text{\}}~\in U^{\text{res}}_s$}
							 {$D^{\text{per}}_{u_\mu,s_\nu} \leftarrow D^{\text{per}}_{u_\mu,s_\nu} \backslash \{d_\lambda\}$} \label{heu_dcpsp_selection:line:kill_datacenter}
					\ENDIF
					\IF{$(V^{\text{res}}_{u_\mu,s_\nu} = 0 ~or~ |D^{\text{per}}_{u_\mu,s_\nu}| = 0 ~or~ l_{LAN} < 1 ~or~ (!ExistLocalCloudlet ~and~ l_{MAN} = 0))$} \label{heu_dcpsp_selection:line:kill_usercluster1}
						\STATE{$U^{\text{res}}_s \leftarrow U^{\text{res}}_s \backslash \{u_\mu,s_\nu\}$} \label{heu_dcpsp_selection:line:kill_usercluster2}
					\ENDIF
				\ENDWHILE
		\ENDFOR 
  \end{algorithmic}
\caption{Heuristic approach for solving the DCPSP}
\label{alg:heu_dcpsp}
\end{algorithm}
%
%
%
\section{\uppercase{Evaluation}}\label{sec:Evaluation}
%
%
\subsection{Setup}
To evaluate our approaches, we prototypically implemented them using Java and IBM CPLEX.
To determine proper test cases, we aim to consider realistic values and parameters.
However, due to the fast-growing number of decision variables when considering multiple time periods, we focus on small problem instances.
We distinguish between (remote) cloud data centers and local cloudlets, \eg, cafes.
As an representation of the \emph{cloud}, we consider one remote cloud data center in our evaluation, along with a variable number of cloudlets.

The two types of data centers are characterized by different capacities, costs, and QoS guarantees.
Based on the commonly agreed characteristics of cloud computing, we assume that the remote cloud data center is able to provide sufficient resources to cover the entire service demand, according to its provided QoS guarantees.
Therefore, we set the maximum capacity $K^{max}_{d_\lambda}$ that is substantially higher than the demand for a specific service.
Furthermore, since we aim to augment the existing infrastructure by local resources, we assume that the cloud resources already exist and, hence, no fixed cost for construction occur.
The provided resources are accessible on-demand when needed and charged in a pay-as-you-go manner.
Hence, we set the minimum capacity constraint $K^{min}_{d_\lambda}$ for the remote cloud data center to \emph{zero}.
Locally installed cloudlets have a limited capacity to host hardware resources, where we assume a maximal capacity between 1 and 20 resource units.
This reflects the obvious association of any deployment with at least one server and the space capacity of a single rack, respectively.

Based on \cite{Greenberg2008Cost}, we assume the following cost ratio for deploying and operating a cloudlet: \emph{(i)} 50\% expenditures for servers, \emph{(ii)} 25\% for fixed infrastructure costs, and \emph{(iii)} 25\% for variable operational costs.
%
Users, cloudlets, and remote data centers are connected through different types of networks: Wide Area Networks (WANs), MANs, and LANs (cf. Figure~\ref{fig:dyn_cloudlet_man}).
For each type of network connection, we assume different download and upload capacities.
Following the published statistics of the New York City Economic Development Corporation (NYCEDC) \cite{NYCEDC2017NYC}, we assume all cloudlet clusters (\eg, cafes), are connected with a symmetric bandwidth of 1 Gbps.
The users' mobile devices within the LAN are connected via Wi-Fi, for which we assume the most recent standard, \ie, IEEE 802.11ac.
For this standard, measurements have shown a realistic bit rate of 500 Mbps up to a distance of 15 meters \cite{Dianu2014Measurement}.
To take full advantage of the given MAN capacity of 1 Gbps, we assume at least two Wi-Fi routers per location.
To represent different store sizes, we set a randomly determined number of Wi-Fi routers between two and six in our simulation.

As described in Section~\ref{sec:problem_statement}, the utilization of cloudlets can be beneficial for a variety of applications.
As discussed, we use three different service classes with different characteristics and costs.
For services with real-time constraints, we assume that cloudlets are strictly required.
The requirement and cost assumptions used for the evaluation are listed in Table \ref{tab:service_req}.
\begin{table}[!h]
\caption{Characteristics of the considered services}
\footnotesize
\renewcommand{\arraystretch}{1.0}
\begin{center}
\begin{tabular} {L{4cm}|L{2.5cm}|L{2.5cm}|L{2.5cm}}
\hline\hline \\[-1em]
	& Service 1 & Service 2 & Service 3\\
\hline
	Migration cost & $1.00 \times C^{mig}_{u_\mu}$ & $0.75 \times C^{mig}_{u_\mu}$ & $0.50 \times C^{mig}_{u_\mu}$\\
	Penalty cost & $1.20 \times C^{pen}_{u_\mu}$ & $1.00 \times C^{pen}_{u_\mu}$ & $1.00 \times C^{pen}_{u_\mu}$ \\
	Req. download cap. (Mbps) & 40 & 40 & 20\\
	Req. upload cap. (Mbps) & 10 & 10 & 20\\
	Latency req. & 50 & 100 & 250\\
	\hline\hline
\end{tabular}
\end{center}
 \label{tab:service_req}
 \end{table}
%
%
%
\subsection{Results and Discussion}
\begin{figure*}[!ht]
	\centering
	\subfloat[Mean computation times (with 95\% CIs). Ordinate in logarithmic scale.\label{subfig:eval_dcpsp_time}]{%
		\includegraphics[width=.5\textwidth]{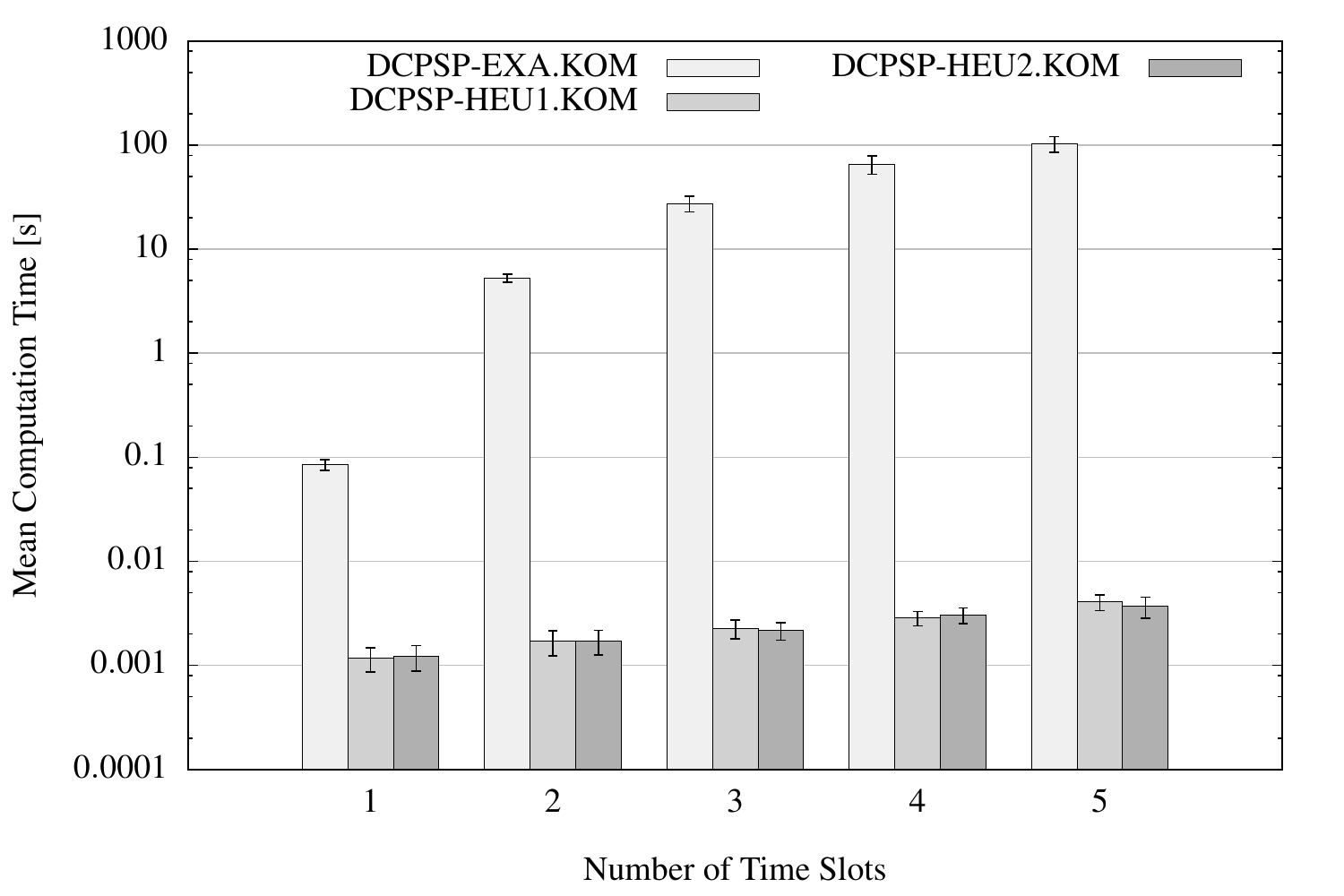}
	}
	\subfloat[Cost ratios (with 95\% CIs).\label{subfig:eval_dcpsp_cost}]{%
		\includegraphics[width=.5\textwidth]{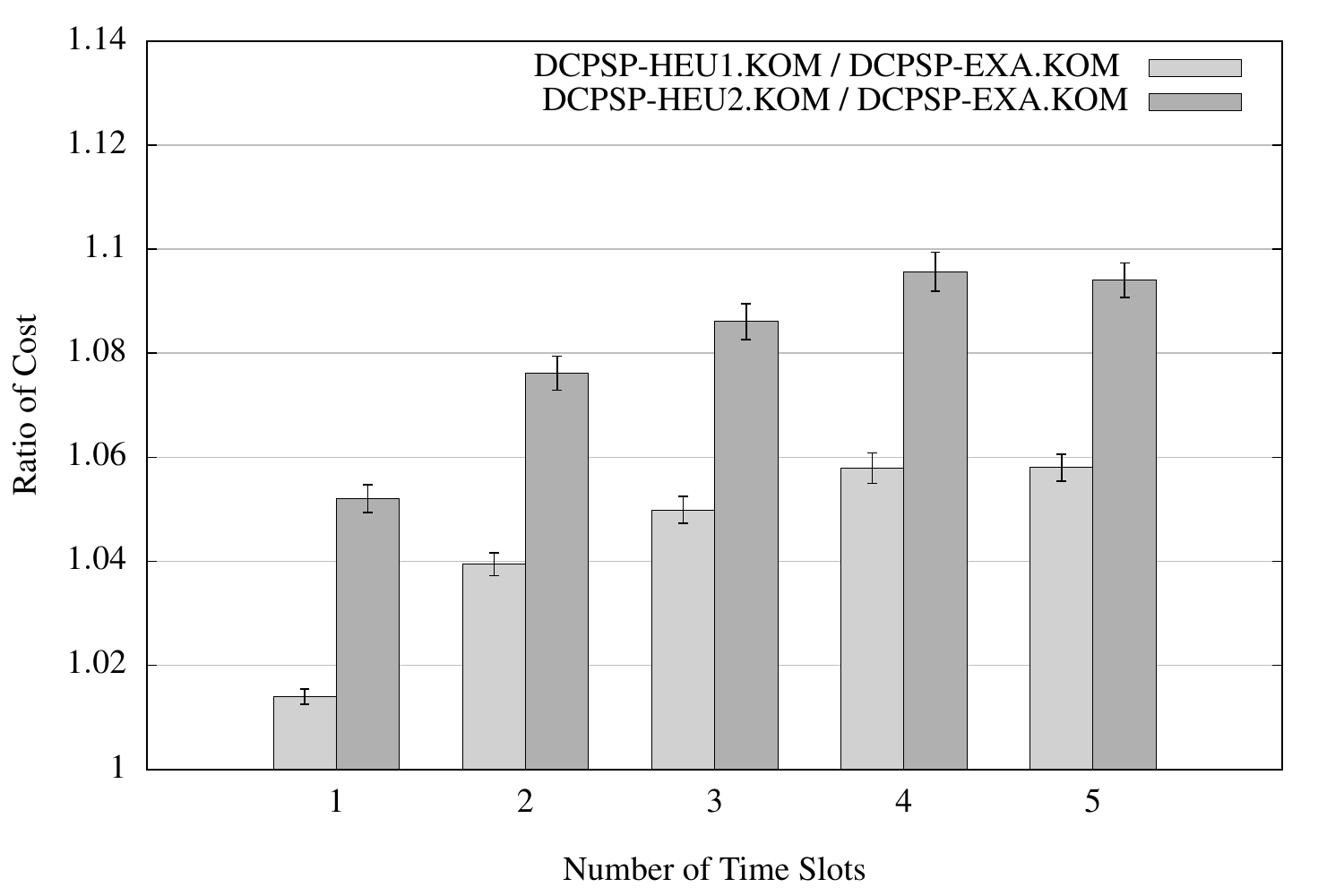}
	}
	\caption{Impact of an increasing number of time slots}
	\label{fig:eval_dcpsp_dc_time}
	\vspace{-10pt}
\end{figure*}

An important factor for our dynamic optimization model are the considered time slots.
While evaluating the number of time slots, all remaining variables are assumed as fixed.
We set the values as follows: $|D| = 20$, $|U| = 20$, $|S| = 3$, and $|Q| = 1$. Due to the high computational effort of the exact solution approach, we limit our simulation to five time slots.
The computation effort of the analyzed approaches are illustrated in Figure~\ref{subfig:eval_dcpsp_time}.

As we expected, the exact approach requires by far the most time to find a solution.
Furthermore, this value increases rapidly with a growing number of considered time slots.
If we consider only one time slot, an average computation time of 84~ms is required. For the last test case with five time slots, the computation time increase by the factor of about $1,230$ to over 103~s.
Both heuristic approaches require nearly the same amount of computation time and show a linear behavior with a growing number of time slots.
The first approach, DCPSP-HEU1.KOM, requires an average computation time of about 1.2~ms for the first test case and 4.1~ms for the last test case.
Hence, compared to the exact approach we are able to reduce the computation time by over \textbf{four orders of magnitude} for the test case with five discrete time slots.

\begin{figure*}[!ht]
	\centering
	\subfloat[Mean computation times (with 95\% CIs). Ordinate in logarithmic scale.\label{subfig:eval_dcpsp_time_dc}]{%
		\includegraphics[width=.5\linewidth]{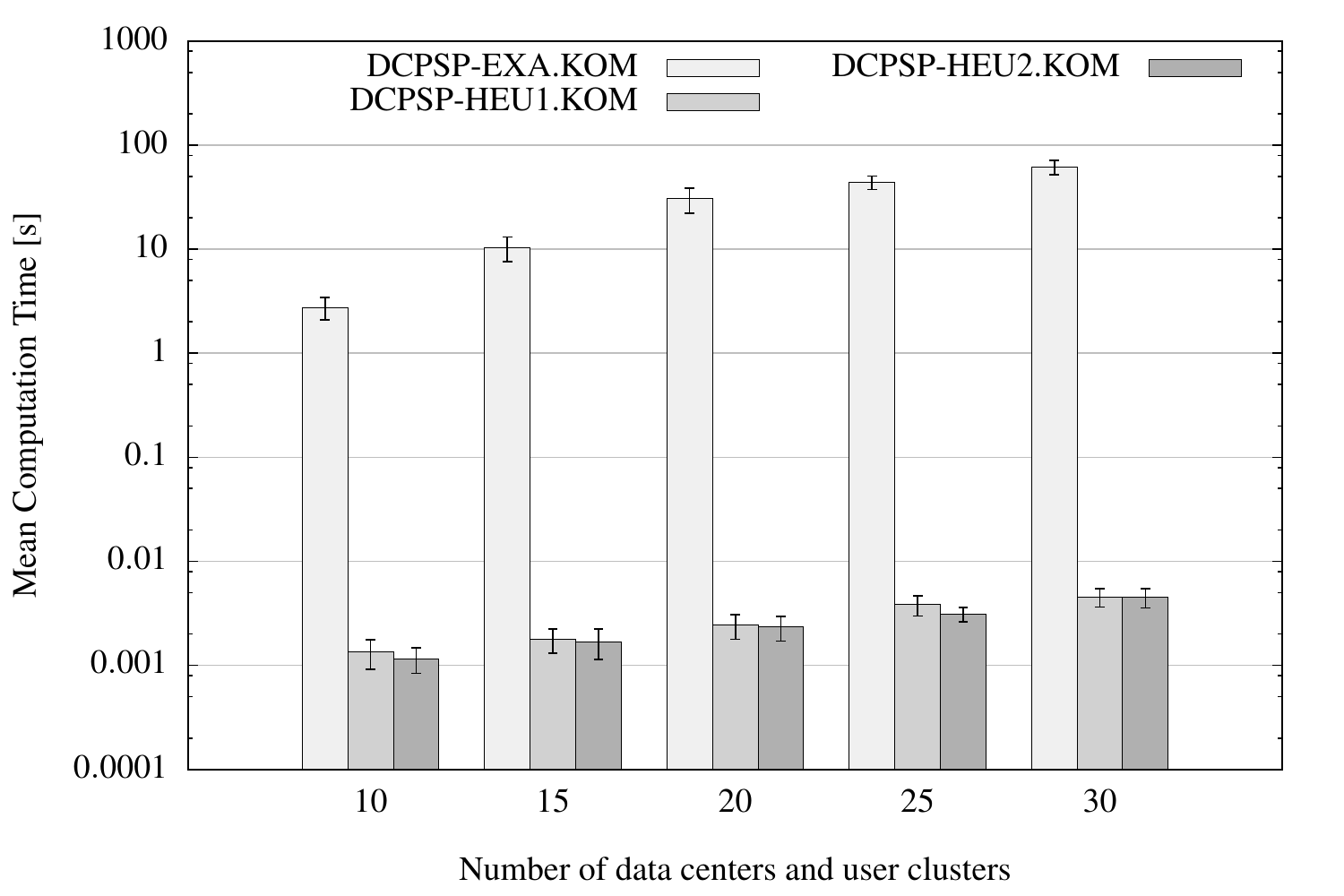}
	}
	\subfloat[Cost ratios (with 95\% CIs).\label{subfig:eval_dcpsp_cost_dc}]{%
		\includegraphics[width=.5\linewidth]{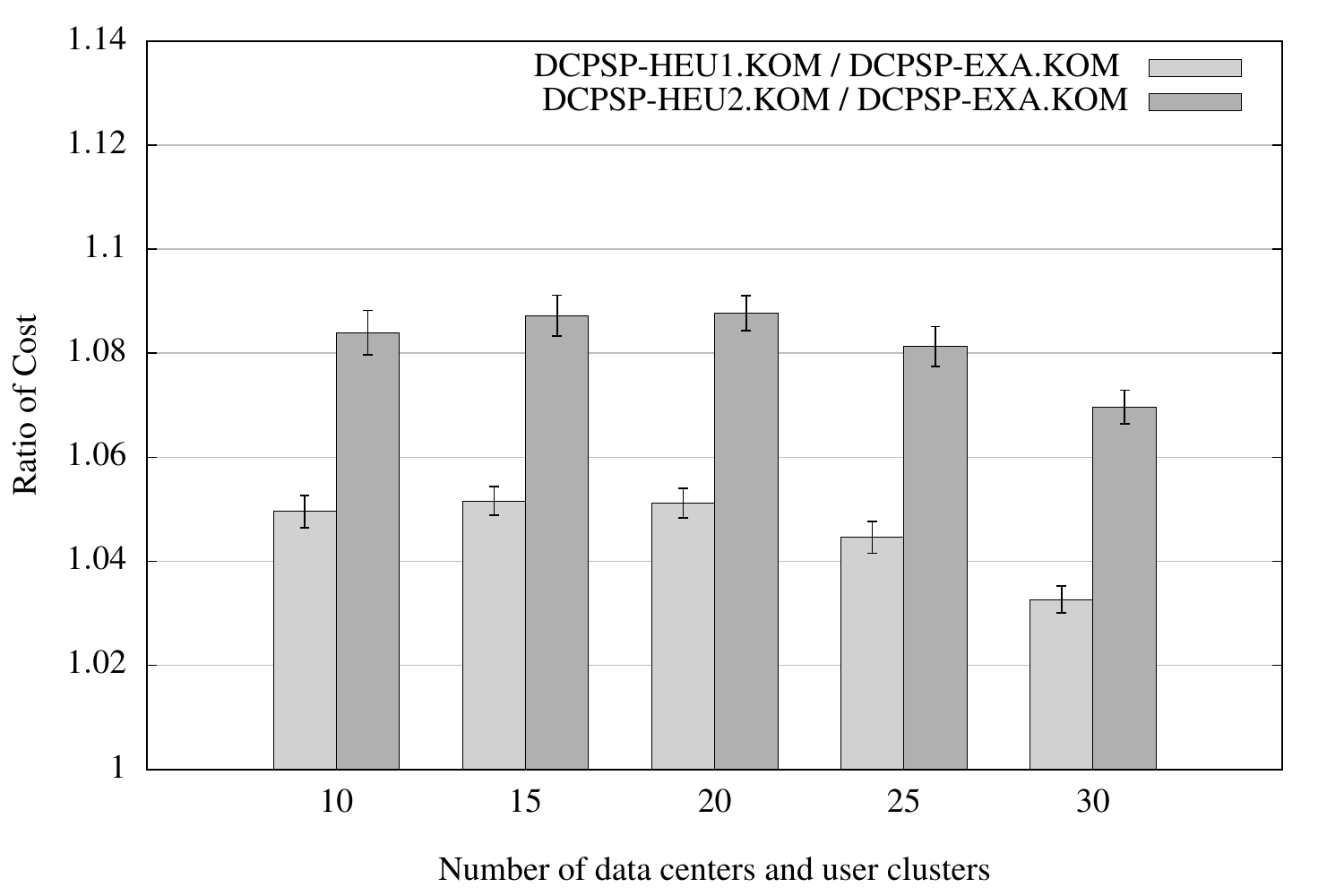}
	}
	\caption{Impact of an increasing number of (local) data centers and user clusters}
	\label{fig:eval_dcpsp_dc_dc}
	\vspace{-10pt}
\end{figure*}

However, the improved computation time comes at the price of a reduced solution quality.
Figure~\ref{subfig:eval_dcpsp_cost} shows the cost ratios of the two strategies.
Two major results are notable: \emph{(i)} The first strategy outperforms the second one by 3.6 to 3.8 percentage points; \emph{(ii)} for both strategies, the solution quality deteriorates with an increasing number of time slots.
However, the solution quality stabilizes from the fourth time slot on.
Focusing on the better approach (DCPSP-HEU1.KOM), it is notable that the solution quality decreases with an increasing number of time slots.
The difference for the first test case corresponds to 1.4\% and increases to 5.8\% for the last test case.
However, the growth of the cost ratio decreases notably between the time slots.
Thus, for the fourth and the fifth time slot, nearly the same solution quality is achieved.
We infer that the reason can be seen in the occurrence of fixed costs.
In a second simulation, we analyzed the influence of data centers and user clusters. Since a cafe or a business location represents a user cluster and can as well act as a cloudlet, we increase both values simultaneously. While analyzing these values, the remaining variables are assumed as constant: $|T| = 3$, $|S| = 3$, and $|Q| = 3$. Figure~\ref{subfig:eval_dcpsp_time_dc} illustrates the computation time depending on the number of considered locations that act as user cluster and local data centers.
Again, the exact solution approach causes the highest computational effort. To find a solution for the first test case ($|D|/|U| = 10$), requires 2.8~s on average. Up the last test case ($|D|/|U| = 30$), this value increases to 61.9~s. Compared to the impact of the number of time slots, the number of data centers/user cluster has a relatively small influence on the computation time. However, using our heuristic approach, we are able to reduce the computation time by at \textbf{least three orders of magnitude}.

The evaluation shows that the limitation of resources is not a promising strategy. Figure~\ref{subfig:eval_dcpsp_cost_dc} shows the cost ratios of our two approaches. Clearly notable are the better results of the first approach. It outperforms the second one by about 3.4 to 3.7 percentage points. An unexpected result is the trend of the solution quality with an increasing number of considered data center/user cluster. For the first case, the difference between the first approach (DCPSP-HEU1.KOM) amounts to 5.0\%. Up to the last test case, the difference shrinks to only 3.3\%. The reason for this result can be seen in a higher number of available resources. Even with an increased resource demand, a higher amount of resources is more able to handle demand fluctuations. Furthermore, with increasing demand and supply, the total costs increase as well and, hence, wrong assignment decisions become less important to the cost ratio.
%
%
%
%
\section{Related Work} \label{sec:relatedwork}
Cloudlets and edge resources are a fast growing research fields. Therefore, in this section we are focusing on two main research areas, \ie, \emph{(i)} Optimization approaches for cloudlet placement, \emph{(ii)} Provisioning of Wi-Fi access nodes in urban areas.

The authors of \cite{Ceselli2015Cloudlet} consider the cloudlet placement within cellular networks.
The authors propose a placement of cloudlets in base stations and formulate a static and a dynamic optimization approach to determine where to install such cloudlets.
In addition to the optimal solution approach, the authors present a heuristic approach.
%
The works in \cite{Dietrich13,Dietrich17} are related to our work as they show relevant problem formulations and corresponding solution heuristics for providing computing resources, as well as, networking resources within a mixed cloud infrastructure network. The main difference to our work is that the authors focus there on the resource provisioning interactions when slicing cloud and connectivity resources across multiple provider networks that do not trust each other.

The authors of \cite{Fesehaye2012Impact} analyze the impact of cloudlets on mobile applications.
The authors present a cloudlet-based peer-to-peer architecture to reduce the propagation delay and increase the throughput.
To evaluate their approach, the authors tested different types of applications, \eg, file editing and video streaming.
With their approach, the authors are able to reduce the transfer delay and the throughput if the maximum number of cloudlet hops does not exceed two.
To handle the rapidly growing mobile traffic, Dimatteo et al. \cite{Dimatteo2011Cellular} present an architecture to evaluate potential costs and gains for providing Wi-Fi offloading.
Therefore the authors simulate the deployment of Wi-Fi hotspots in San Francisco and aim to determine how many access points are required to obtain a performance improvement.
The authors use a ranked list and freely choose a location within the metropolitan area.
In our work, we are tied by a set of possible locations, since we use stationary cafes as cloudlet locations and user clusters.
Further, in contrast to the related work that tries to improve the download delay, we are focusing on highly interactive applications and thus, on latency.
%
%
\section{Conclusion}
\label{sec:conclusion}
To provide services with stringent QoS requirements, an augmentation of the centralized cloud infrastructure by locally installed cloudlets is a promising approach. Since the utilization of decentralized micro data center is costly, we examined the \emph{Dynamic Cloudlet Placement and Selection Problem} to provide the means of a cost-efficient infrastructure augmentation. We formulate the mixed integer optimization problem to compute the exact solution. To overcome the problem of high computational effort, we investigate different heuristic approaches. By the means of these heuristics, we are able to significantly reduce the computation time by at least three orders of magnitude while maintaining a high solution quality, with a moderate increase of cost by up to 5.8\%.
%
%
\section*{Acknowledgment}
This work has been sponsored in part by the German Federal Ministry of Education and Research (BMBF) under grant no. 01IS12054, by E-Finance Lab e.V., Frankfurt a.M., Germany (www.efinancelab.de), and by the German Research Foundation (DFG) in the Collaborative Research Center (SFB) 1053 – MAKI.
%
%
%
\bibliographystyle{splncs04}
\bibliography{ESOCC}
\end{document}